\documentclass[12pt]{article}
\usepackage{amsmath,color,epsfig,amsfonts,fullpage,natbib,amssymb,multirow,multicol,float,setspace}
\usepackage[titletoc,title]{appendix}
\usepackage{epstopdf}
\newcommand{\vv}[1]{\mbox{\boldmath $#1$}}

\newcommand{\beq}{\begin{eqnarray*}}\newcommand{\eeq}{\end{eqnarray*}}
\newcommand{\beqn}{\begin{eqnarray}}\newcommand{\eeqn}{\end{eqnarray}}
\newcommand{\bit}{\begin{itemize}}\newcommand{\eit}{\end{itemize}}
\newcommand{\bnum}{\begin{enumerate}}\newcommand{\enum}{\end{enumerate}}

\newcommand{\tran}{^\intercal}
\newcommand{\inv}{^{-1}}

\newcommand{\btheta}{\boldsymbol\theta}

\newcommand{\comment}[1]{}
\makeatletter
\newcommand{\thickhline}{%
    \noalign {\ifnum 0=`}\fi \hrule height 2pt
    \futurelet \reserved@a \@xhline
}

\makeatother

\DeclareMathOperator*{\argmax}{arg\,max}   % In your preamble
  
\begin{document}
%% ======================================================================

\begin{center}
{\setstretch{1.51
}\section*{Maximum likelihood estimation for\\ stochastic differential
  equations using\\ sequential kriging-based optimization}}
\vspace{-.45cm}

\textbf{Grant Schneider}$^{1,3}$,
\textbf{Peter F. Craigmile}$^{1,2,4}$ and
\textbf{Radu Herbei}$^{1,5}$\\[.1cm]
\vspace{-.25cm}

$^1$
Department of Statistics,
The Ohio State University,
Columbus, OH 43210, USA
\vspace{-.25cm}

$^2$
School of Mathematics and Statistics,
University of Glasgow,
Glasgow,
Scotland
\vspace{-.55cm}
$^3$\verb_schneider.393@osu.edu_ \;
$^4$\verb_pfc@stat.osu.edu_ \;
$^5$\verb_herbei@stat.osu.edu_
\end{center}

\vspace{.1cm}

\textbf{Abstract:$\;$} 
Stochastic Differential Equations (SDEs) are used as statistical models in many disciplines.  
However, intractable likelihood functions for SDEs make inference challenging, and we need to resort to simulation-based techniques to estimate and maximize the likelihood function.  
While sequential Monte Carlo methods have allowed for the accurate
evaluation of likelihoods at fixed parameter values, there is still a
question of how to find the maximum likelihood estimate.
In this article we propose an efficient Gaussian-process-based method
for exploring the parameter space using estimates of the likelihood
from a sequential Monte Carlo sampler.
Our method accounts for the inherent Monte Carlo variability of the
estimated likelihood, and does not require knowledge of gradients.
The procedure adds potential parameter values by maximizing the
so-called expected improvement, leveraging the fact that the
likelihood function is assumed to be smooth.
%
%Confidence regions for the parameters can also be obtained. 
%
Our simulations demonstrate that our method has significant
computational and efficiency gains over existing grid- and
gradient-based techniques.
Our method is applied to modeling the closing stock price of three technology firms.

\textbf{Keywords:$\;$}
Discretely sampled diffusions;
Expected improvement;
Gaussian process;\\
Sequential Monte Carlo;
Parameter estimation.

\section{Introduction}
\label{sec:introduction}
Many phenomena that arise in finance, biology, ecology, and other areas are modeled in continuous time using a real-valued diffusion process, $\{X_t\}$, that is the solution to the \emph{stochastic differential equation} (SDE)
\begin{align}
\label{diffusion}   
    dX_t=\mu(X_t,\boldsymbol\theta) \,
    dt+\sigma(X_t,\boldsymbol\theta) \, dW_t, \quad 0\le t \le T,
\end{align}
where $X_0=x_0$ is the initial value of the process and $\{W_t\}$ is a standard Brownian motion. We assume that the drift and the diffusion functions, $\mu( \cdot,\cdot)$ and $\sigma( \cdot,\cdot )$ respectively, are known up to the parameter vector $\boldsymbol\theta\in\boldsymbol\Theta$, where $\boldsymbol\Theta$ is some compact set in $\mathbb{R}^p$. We further assume the drift and diffusion functions are locally Lipschitz with linear growth bounds so that a weakly unique solution to \eqref{diffusion} exists.
Suppose that we observe the process $\{X_t\}$ at time points $t_i$ $(i=1,\dots,N)$ where $0=t_0<t_1<\dots<t_N$, and let $\mathbf{X}=(X_{t_1},\dots,X_{t_N})\tran$. In this article we are interested in maximum likelihood estimation of $\boldsymbol\theta$ and associated confidence bounds based on the data $\mathbf{X}$. 

Let $p(x|x_{t_{i-1}},\boldsymbol\theta)$ represent the conditional probability density of $X_{t_i}$ given $X_{t_{i-1}}=x_{t_{i-1}}$ evaluated at $x$ for a given set of parameters $\boldsymbol\theta\in\boldsymbol\Theta$. Treating $X_0=x_0$ as fixed, we can then use the Markov property to write the likelihood of the data as the product of these individual \emph{transition densities}
\begin{align}
L(\boldsymbol\theta|\mathbf{X})&=\prod\limits_{i=1}^{N}p(X_{t_i}|X_{t_{i-1}},\boldsymbol\theta).
\label{likelihood}
\end{align}

When the transition density is known, likelihood calculation and its
maximization with respect to $\boldsymbol\theta\in\boldsymbol\Theta$
for a given set of data discretely observed from \eqref{diffusion} is
straightforward. As the transition density does not exist in
closed-form except for a handful of cases, approximations are
typically necessary; see for example \cite{hurn2007seeing} for a
recent overview. These methods are often separated into four groups:
(1) sequential Monte Carlo (SMC)
\citep{Pedersen1995,santa1997simulated,elerian2001likelihood,brandt2002simulated,DurhamGallant2002,
  lin2010generating}, (2) methods based on the exact simulation of
diffusions \citep{EAEstimate}, (3) closed-form Hermite expansions of
the transition density \citep{AitSahalia,ait2008closed}, and (4)
approximations derived by numerically solving the Kolmogorov forward
equation \citep{lo1988maximum}.

Detailed discussions of the benefits of the method on which we choose
to focus, SMC, may be found in
\cite{DurhamGallant2002,brandt2002numerical,brandt2002simulated} and a
direct comparison to the Hermite expansion in
\cite{StramerYan2007}. In contrast to the other procedures, it does
not require transforming \eqref{diffusion} into an SDE of unit
diffusion and it can be made arbitrarily accurate at the expense of
more computation. One of our goals is to ease this computational
burden associated with repeatedly obtaining Monte Carlo estimates of
the log-likelihood over the parameter space. Much of the previous work
has focused on efficiently estimating the likelihood at a fixed
$\boldsymbol\theta$. An approximate MLE can then be obtained by
maximizing this estimated likelihood over $\boldsymbol\Theta$. In some
cases, the derivatives of the log-likelihood with respect to
$\boldsymbol\theta$ can be obtained from the simulated values used to
produce the estimate of the likelihood as in \cite{StramerYan2007} and
gradient ascent optimization is straightforward. Typically, however,
these derivatives must be obtained numerically, which adds a
significant computational burden. Although the underlying
log-likelihood may be smooth as a function of $\boldsymbol\theta$, the
Monte Carlo estimates will be subject to variability and thus will be
much less amenable to derivative calculation.

We propose an efficient Gaussian-process-based method for exploring
$\boldsymbol\Theta$ which accounts for the inherent Monte Carlo
variability of the simulated likelihood method and does not require
knowledge of the gradient of the log-likelihood. Our sequential method
offers significant computational efficiency over a naive approach
based on estimating the likelihood over a grid of possible parameter
values. By using a global search criterion from computer experiments
called the ``expected improvement''
\citep{jones1998efficient,schonlau1998global,WilliamsSantnerNotz}, we
alleviate difficulties with local maxima that may be encountered by
gradient ascent methods.  We obtain a sequential sample of parameter
values, thus avoiding the need to use a regular grid for likelihood
evaluation, through a kriging approach which allows for estimation of
the MLE of $\btheta$ and straightforward quantification of its
uncertainty.

The paper is organized as follows. Section \ref{sec:background} introduces and discusses
the SMC method for estimating the likelihood at fixed parameter values
$\boldsymbol\theta$. Section \ref{sec:kboptim} details our proposed sequential
kriging-based optimization method that is used to find the MLE of
$\btheta$ and to construct confidence bounds for $\btheta$. In Section
\ref{sec:simulation} we evaluate the performance of our MLE method through a simulation
study. We apply our methodology to the modeling of the closing price of three technology stocks in Section \ref{sec:data}.  Conclusions and discussion are given in Section \ref{sec:discussion}. A discussion of our choice of importance sampler is given in Appendix \ref{sec:modBB}. 

\section{Sequential Monte Carlo}
\label{sec:background}

Central to our approach is the ability to approximate the transition density $p(X_{t_i}|X_{t_{i-1}},\boldsymbol\theta)$, for each $i=1,\ldots,N$.  Without loss of generality, it is sufficient to approximate $p(X_{\Delta}|X_{0},\boldsymbol\theta)$ for some $\Delta > 0$.
We use the Euler approximation given by
\begin{align}\label{Euler}
p\left(X_\Delta|X_0,\btheta)\approx     \xi(X_{\Delta}|X_0,\boldsymbol\theta\right)
=
     \phi\left(X_{\Delta};X_0+\mu(X_0,\boldsymbol\theta)\Delta,\sigma^2(X_0,\boldsymbol\theta)\Delta\right),
\end{align}
where $\phi\left(x;m,\nu^2\right)$ is the density of a normal random variable with mean $m$ and variance $\nu^2$ evaluated at $x$.
In general, for large $\Delta$ this approximation is inaccurate. As a remedy, we partition the interval $[0,\Delta)$ into $K$ subintervals with endpoints $0=\tau_0<\tau_1<\dots<\tau_K=\Delta$, such that $\tau_j-\tau_{j-1} = \Delta/K$, $j=1,2,\dots,K$ and consider the unobserved points $\mathbf{X}_\tau=\left(X_{\tau_1},X_{\tau_2},\dots,X_{\tau_{K-1}}\right)$. The \emph{discretized} transition density \cite[see][]{kloeden1992numerical} is defined to be
\begin{align}
\label{discTran}
    p^{(K)}\left(X_\Delta|X_0,\boldsymbol\theta\right)&=\int \prod\limits_{k=1}^K 
\xi\left(X_{\tau_k}|X_{\tau_{k-1}},\boldsymbol\theta\right)
\lambda\left(d\mathbf{X}_\tau\right),
\end{align}
where $\lambda$ denotes the Lebesgue measure. We can use importance sampling to calculate \eqref{discTran} by calculating the expectation of the random variable $R^{(K)}$ with respect to the \emph{importance density} $q(\cdot)$, where $R^{(K)}={\prod\limits_{k=1}^K \xi\left(X_{\tau_k}|X_{\tau_{k-1}},\boldsymbol\theta\right)}/{q\left(\mathbf{X}_\tau\right)}$.

We use a classical Monte Carlo estimator for this expectation, given by
\begin{align}
\label{MCTran}
p^{(K,M)}\left(X_\Delta|X_0,\boldsymbol\theta\right)=\frac{1}{M}\sum\limits_{m=1}^M R^{(K)}_m.
\end{align}
where the variates $R^{(K)}_m$ $(m=1,\dots,M)$ are calculated based on $M$ independent and identically distributed (IID) draws $\mathbf{X}_{\tau,m}=\left(X_{\tau_1,m},\dots,X_{\tau_{K-1},m}\right)\sim q(\cdot)$. The construction of the discretized and estimated log-likelihoods is a straightforward application of the Markov property as in \eqref{likelihood}.

It is clear from \eqref{discTran} and \eqref{MCTran} that selecting a good importance sampling density $q(\cdot)$ is very important. The optimal $q(\cdot)$ is the true joint density of $\mathbf{X}_\tau$ given $X_0$ and $X_\Delta$ \citep{StramerYan2007}, which is unavailable, as it depends on the transition density that we are trying to estimate.
\cite{Pedersen1995} and \cite{brandt2002simulated} choose $q\left(\mathbf{X}_\tau\right)=\prod\limits_{k=1}^{K-1} \xi\left(X_{\tau_k}|X_{\tau_{k-1}},\boldsymbol\theta\right)$. In that case, $R^{(K)}$ simplifies to $\xi\left(X_\Delta|X_{\tau_{K-1}},\boldsymbol\theta\right)$ and \eqref{MCTran} can be rewritten as
\begin{align}
\label{pedersen}
p^{(K,M)}\left(X_\Delta|X_0,\boldsymbol\theta\right)=\frac{1}{M}\sum\limits_{m=1}^M \xi\left(X_\Delta|X_{\tau_{K-1},m},\boldsymbol\theta\right).
\end{align}
We note that simulating $\mathbf{X}_{\tau,m}\sim q(\cdot)$ follows from sequentially simulating $X_{\tau_{j},m}|X_{\tau_{j-1},m},\btheta$, for $j=1,\dots,K-1$ according to \eqref{Euler}. \cite{elerian2001likelihood} criticized \eqref{pedersen} for its inefficiency and proposed a computationally intensive method of sampling $\mathbf{X}_\tau$ from a multivariate Normal or $t$ distribution based on a second-order Taylor expansion. Exact simulation \citep{EA1,EA2,EA3} presents an opportunity to sample from a $q(\cdot)$ that is exactly $p\left(\mathbf{X}_\tau|X_{0},\boldsymbol\theta\right)$, but adds another layer of computational complexity \citep{bladt2014simple}. Similar to Pedersen's method \eqref{pedersen}, the exact simulation method ignores $X_\Delta$, and thus suffers from efficiency issues.

As a compromise between the accuracy and the computational efficiency of estimating the log-likelihood, we choose $q(\cdot)$ to be the modified Brownian bridge sampler \citep{DurhamGallant2002}. This sampler produces values of $\mathbf{X}_\tau$ conditional on $X_0$ and $X_\Delta$, which leads to an efficiency gain over Pedersen's method and the exact simulation techniques. Further details of this sampler are given in Appendix \ref{sec:modBB}. An added benefit of using the modified Brownian bridge sampler is that guidance is available for the somewhat arbitrary choices of $K$ and $M$. We set $M=K^2$, as \cite{StramerYan2007b} show that this choice is computationally optimal for a fixed large amount of computer time. A recent extension, the guided-resampling version of the Brownian bridge sampler \citep{lin2010generating}, may be used when $\sigma(\cdot)$ depends strongly on $X_t$ or $\Delta$ is large. As the original version of the modified Brownian bridge sampler is sufficient for our purposes, we avoid the bit of extra computation and set-up costs associated with the guided-resampling version.

We emphasize that although we choose to use the modified Brownian bridge sampler, the user is free to choose any $q(\cdot)$ desired before proceeding with our proposed method. The references above provide guidance for obtaining estimates at a fixed $\btheta$, but exploring the parameter space remains problematic. Although a very important practical issue, exploration methods have been typically ignored in the literature.  A notable exception is in \cite{lin2010generating}, where the authors admit that their rough estimate of the smooth likelihood is not conducive to parameter estimation. 

\section{Kriging-Based Optimization}
\label{sec:kboptim}

Our approach assumes that the discretized log-likelihood function,
$l^{(K)}\left(\btheta\right)$, is smooth in $\btheta$, but that the estimates of
this function obtained via the Brownian bridge sampler (our SMC
estimate) are subject to Monte Carlo variability. Rather than
attempting to maximize the estimated function using a prohibitively
large Monte Carlo sample, we propose a sequential optimization method
that explicitly models the underlying smooth discretized
log-likelihood using a Gaussian process (GP), while treating Monte
Carlo variability as measurement error. Using kriging equations,
parameter values are added sequentially by maximizing the so-called
expected improvement, which balances the uncertainty in estimating the
discretized log-likelihood at unexplored parameter values with the
desire to find parameter values near the current maximum that have a
higher log-likelihood.

More formally, given data $\mathbf{X}$ from \eqref{diffusion}, we
start by estimating the discretized log-likelihood
$l^{(K)}\left(\btheta\right)=\sum\limits_{i=1}^N\log
p^{(K)}\left(X_{t_i};X_{t_{i-1}},\btheta\right)$ at a range of parameters values,
$\btheta$, that span the space of possible parameter values.  For this
purpose let $\left(\btheta_1, \ldots, \btheta_n\right)\tran$ denote the initial $n$
parameter values, selected using a space-filling design, such as a
Latin hypercube.
Letting $Y\left(\btheta_i\right)$ denote the SMC-based estimate of
$l^{(K)}\left(\btheta_i\right)$, we assume for $i=1,\ldots,n$ that
\beqn\label{eq:datamodel}
    Y\left(\btheta_i\right) = l^{(K)}\left(\btheta_i\right) + \epsilon\left(\btheta_i\right),
\eeqn
where $\left\{ \epsilon\left(\btheta_i\right) : i=1,\ldots,n \right\}$ is a set of independent
of N($0$, $\sigma^2$) errors. 

We model $\{l^{(K)}\left(\boldsymbol\theta\right) : \boldsymbol\theta\in
{\boldsymbol\Theta}\}$ using a  GP with mean function $\mu_L\left(\btheta;\beta\right)$ and some valid
covariance function $C_L\left(\btheta,\btheta';\zeta\right)$, where $\beta, \zeta$ are
unknown parameters.  There is an extensive literature on the choice of mean and
covariance function for the GP  \cite[see][]{cressie2011statistics,santner2003design}
-- richer choices can more accurately emulate the discretized log-likelihood, at a cost of 
necessitating larger sample sizes, $n$, and more computational resources to
faithfully model the features of the GP.

The sequential optimization procedure proceeds as follows. Given $\vv{Y}_n = \left( Y(\btheta_1), \ldots,
Y(\btheta_n) \right)\tran$, we first find estimates $\widehat
\beta_n,\widehat\zeta_n$ of  the parameters defining the GP above. Conditionally on
$\vv{Y}_n, \widehat
\beta_n,\widehat\zeta_n$, we predict  the
discretized log-likelihood $l^{(K)}\left(\btheta^*\right)$ for some
  $\btheta^*\in \boldsymbol\Theta$.  By Gaussianity of the GP and data model
\eqref{eq:datamodel}, the best linear unbiased prediction of
$l^{(K)}\left(\btheta^*\right)$ is the kriging mean
\beqn
\label{eq:kriging:mean}
    \eta_L\left(\btheta^*\right)
&=&
    \mu_L\left(\btheta^*\right) +
    \vv{c}_L \tran ( \vv{\Sigma}_L + \sigma^2 \vv{I}_n )^{-1} 
    [ \vv{Y}_n - \vv{\mu}_L ],
\eeqn
with kriging variance
\beqn
\label{eq:kriging:var}
    v^2_L\left(\btheta^*\right)
&=&
    C_L(\btheta^*, \btheta^*) -
    \vv{c}_L \tran ( \vv{\Sigma}_L + \sigma^2 \vv{I}_n )^{-1} \vv{c}_L.
\eeqn
In the above equations $\vv{\mu}_L$ is a mean vector of length $n$
with $i$th element $\mu_L(\btheta_i;\widehat \beta_n)$, $\vv{c}_L$ is a covariance
vector of length $n$ with $i$th element $C_L(\btheta^*, \btheta_i;\widehat\zeta_n)$, and
$\vv{\Sigma}_L$ is the $n \times n$ covariance matrix with $(i,j)$
element $C_L(\btheta_i, \btheta_j;\widehat \zeta_n)$.
Let $\widetilde{\eta}_L = \max_{i=1,\ldots,n} \eta_L\left(\btheta_i\right)$
denote the maximum value of the kriging mean over the explored $\btheta$ values.  Then, the
\textit{improvement}~\citep{jones1998efficient} at $\btheta^*$ is
\beqn
    I\left(\btheta^*\right)
&=&
    \max\left\{ 0, l^{(K)}\left(\btheta^*\right) - \widetilde{\eta}_L \right\},
\eeqn
but since $l^{(K)}\left(\btheta^*\right)$ is unknown we replace it by the \textit{expected improvement} at parameter value $\btheta^*$, which can be shown to be equal to~\citep{jones1998efficient}
\beqn
\label{expected:improvement}
    E( I\left(\btheta^*\right) | \vv{Y}_n )
&=&
    \left[ \eta_{L}\left(\btheta^*\right)-\widetilde{\eta}_L \right]
    \Phi\!\left(\frac{\eta_{L}\left(\btheta^*\right)-\widetilde{\eta}_L}{v_{L}\left(\btheta^*\right)} \right) +
    v_{L}\left(\btheta^*\right) \;
    \phi\!\left(\frac{\eta_{L}\left(\btheta^*\right)-\widetilde{\eta}_L}{v_{L}\left(\btheta^*\right)}
    \right),
\eeqn
where $\Phi(\cdot)$ is the standard Gaussian cumulative distribution function.
As explained above, the expected improvement balances the need
to maximize the discretized log-likelihood (the first term) and the
uncertainty in estimating the log-likelihood (the second term).

Our sequential optimization scheme adds the parameter value $\btheta^*$
that maximizes the expected improvement \eqref{expected:improvement};
 details are given at the end of this section.
We then estimate the discretized log-likelihood using SMC at
that new parameter value, yielding $Y\left(\btheta^*\right)$  and update the data $\vv{Y}_{n+1} =
\left(\vv{Y}_n\tran, Y\left(\btheta^*\right)\right)\tran$ and the GP parameter estimates
$\widehat \beta_{n+1}, \widehat\zeta_{n+1}$.
After updating the kriging mean \eqref{eq:kriging:mean} and kriging
variance \eqref{eq:kriging:var} using $\vv{Y}_{n+1}, \widehat \beta_{n+1}, \widehat\zeta_{n+1}$, we search for another
parameter value that maximizes the expected improvement.  We continue
observing new estimated log-likelihood values and optimizing
\eqref{expected:improvement} to find more parameter values, until some
stopping criteria is met.
It is then straightforward to obtain the estimated MLE,
$\widehat{\btheta}=\argmax\limits_{i=1,\dots,n}\eta_L\left(\btheta_i\right)$,
where $n$ is the total number of iterations in this procedure.

We can also obtain an approximate $(1-\alpha)\%$ joint confidence region for $\btheta$ directly from the kriging mean based on the likelihood ratio test:
\begin{align}
\label{likelihoodratioCI}
\left\{\btheta: 2\left(\eta_L(\widehat{\btheta})-\eta_L\left(\btheta\right)\right)\le \chi^2_{1-\alpha,p}\right\},
\end{align}
where $\chi^2_{1-\alpha,p}$ is the 1-$\alpha$ quantile of a chi-square distribution with $p$ degrees of freedom.
Given the GP parameters $\eta_L(\cdot)$ will be will be trivial to compute compared to $Y(\cdot)$. 
However as the dimension of $\btheta$, $p$, grows obtaining the region in \eqref{likelihoodratioCI} may become difficult;
in that case we can use a Rao-based confidence region of the form
%
%When \eqref{likelihoodratioCI} proves cumbersome to construct, we can instead obtain the confidence region given by
%we can proceed by recalling the Fisher information matrix $\mathcal{I}(\btheta)$ with entry $i,j$ given by $-E\Big[l^{\prime\prime}(\btheta|\X)\Big|\btheta\Big]$, where $l^{\prime\prime}(\btheta|\X)$ is the second derivative of $l(\btheta|\X)$ with respect to $\btheta$. 
%
%Using the fact that asymptotically, $\widehat{\btheta}\sim N(\btheta, \mathcal{I}(\btheta)\inv)$, and estimating $\mathcal{I}(\theta)$ with $\widehat{\mathcal{I}}(\widehat{\btheta})=\left. \eta^{\prime\prime}_L({\btheta})\right|_{\btheta=\widehat{\btheta}}$, we can construct an approximate $(1-\alpha)\%$ joint confidence region for $\btheta$ given by
\begin{align}
\label{WaldCI}
\left\{\btheta: \left(\widehat{\btheta}-{\btheta}\right)\tran{\widehat{\mathcal{I}}(\widehat{\btheta})}\inv\left(\widehat{\btheta}-{\btheta}\right)\le \chi^2_{1-\alpha,p}\right\},
\end{align}
where we can estimate the Fisher information, $\widehat{\mathcal{I}}(\widehat{\btheta})$ using the second derivative of $\eta_L(\cdot)$ with respect to $\btheta$ evaluated at $\widehat{\btheta}$.

\paragraph{Estimating the parameters of the GP.}

 We now turn to estimating the parameters $\beta$ and $\zeta$
  conditionally on $\vv{Y}_n$. These parameters may be updated in a
  variety of ways, including least squares, maximum likelihood,
  restricted maximum likelihood, and Bayesian methods. We adopt a
  Bayesian viewpoint and specify a  prior distribution $[\beta,
  \zeta]$ for the GP parameters.  Throughout the paper we use
  the square bracket notation $[\cdot]$ to denote ``distribution
  of''. For notational simplicity,
  let 
${\rm E}(\vv{Y}_n\mid \beta, \zeta) = \mu_Y$ and ${\rm Var}(\vv{Y}_n \mid \beta,\zeta) = \Sigma_Y$.
Using Bayes' rule,
\begin{align}
\label{eq:post}
[\beta,\zeta\mid \vv{Y}_n] &\propto
[\beta,\zeta][\vv{Y}_n\mid\beta,\zeta] \nonumber \\ 
&=[\beta,\zeta]|\Sigma_Y|^{-1/2}\exp\left\{ -\frac{1}{2} ( \vv{Y}_n -
  \mu_Y )\tran \Sigma_Y^{-1} (\vv{Y}_n - \mu_Y) \right\} \
\end{align}
The posterior distribution \eqref{eq:post} can be explored and
summarized in several ways. In our examples below, we take our estimates
$\big(\widehat \beta_n,\widehat \zeta_n\big)$ to be the mode of \eqref{eq:post}.

%Adapting \cite{WilliamsSantnerNotz}, we have
%\begin{align}
%\label{posteriorzeta}
%[\zeta|\Y]\propto[\zeta]\frac{|\Sigma_Y|^{-1/2}}{\sqrt{\1\tran\SY\inv\1}}\exp\Big\{-\frac{1}{2}\bigg(\Y\tran\SY\inv\Y-\frac{\big(\1\tran\SY\inv\Y\big)^2}{\1\tran\SY\inv\1}\bigg)
%\Big\}
%\end{align}
%
%Although a fully Bayesian approach could be taken by integrating $\zeta$ out of \eqref{posteriorzeta}, we follow \cite{WilliamsSantnerNotz} and update the covariance parameter values to be equal to their posterior modes, which are the restricted maximum likelihood (REML) estimators \citep{Cres:1993}.
%
%Once we have obtained the covariance parameters and calculated $\Sigma_Y$, we can calculate the MLE for $\beta$ to be
%\begin{align}
%\widehat{\beta}&=(\1\tran\Sigma_Y\inv\1)\inv\1\tran\Sigma_Y\inv \y.
%\end{align}

\paragraph{Maximizing the expected improvement.}
\label{sec:max}
Maximizing \eqref{expected:improvement} may be done using either an optimization procedure (e.g. Nelder-Mead) or over a grid. 
If an optimization approach is used, to aid the exploration of $E( I(\cdot) | \vv{Y}_n )$ over $\btheta$, the derivative of \eqref{expected:improvement} with respect to $\btheta^*$ can be shown to be
\begin{align}
\frac{d}{d\btheta^*} E( I\left(\btheta^*\right) | \vv{Y}_n )
&=
\frac{d\eta_L\left(\btheta^*\right)}{d\btheta^*}  \Phi\!\left(\frac{\eta_{L}\left(\btheta^*\right)-\widetilde{\eta}_L}{v_{L}\left(\btheta^*\right)} \right)+
\frac{dv_L\left(\btheta^*\right)}{d\btheta^*}\phi\!\left(\frac{\eta_{L}\left(\btheta^*\right)-\widetilde{\eta}_L}{v_{L}\left(\btheta^*\right)} \right).
\end{align}
The derivatives $d\eta_L\left(\btheta^*\right)/{d\btheta^*}$ and ${dv_L\left(\btheta^*\right)}/{d\btheta^*}$ will depend on the mean and covariance functions of the chosen GP and may be calculated accordingly.
The grid approach may be simpler to implement, but it should be noted
that this will become computationally intensive as the dimension of
$\btheta$ grows.
This will be similar to the problem with estimating $l^{(K)}(\cdot)$ over a fine grid, but it will occur at a slower rate because the computations involved in calculating expected improvement are simpler than those used to estimate the discretized log-likelihood.
If a grid approach is used, adjustments should also be made to allow for replicates at a given~$\btheta_i$.

\paragraph{Choosing a stopping rule.}
We choose to stop adding points once we have observed no change to $\widehat{\btheta}$ for five consecutive added points.
As noted in \cite{WilliamsSantnerNotz}, expected improvement is not monotonically decreasing as points are added, so choosing a stopping rule may prove difficult.
Possible alternative choices for stopping criteria may be based on a fixed number of points $n$ or small changes in $\eta_L(\cdot)$. Alternatively, consecutively observing the maximum expected improvement over $\boldsymbol\Theta$ below some threshold is a viable strategy as~well.

\section{Simulation Results}
\label{sec:simulation}

We now use data simulated from two models to evaluate the performance of the sequential kriging-based optimization (SKBO) compared to naive space-filling designs in terms of accuracy and speed. 
We consider a ``practical'' naive space-filling design which estimates $l^{(K)}(\cdot)$ at $25p$ points across $\boldsymbol\Theta$. 
We choose to stop the sequential search when the estimate of $\widehat{\btheta}$ changes by less than .01 in each direction for five consecutive iterations, or when we have sampled $25p$ points, whichever occurs first, to provide a fair comparison to the naive~method.

A common rule-of-thumb in the computer experiments literature is to use $n=10p$ initial $Y(\cdot)$ values based on a space-filling design. 
To investigate the role that the initial number of points plays, we perform SKBO based on $n=5p$ and $n=10p$ initial $Y(\cdot)$ values. 
We also attempted to investigate the alternative strategy of using a large $(M=25000)$ Monte Carlo sample to guide a steepest-ascent search, but even at this value of $M$, in which the estimated $Y(\cdot)$ values took 1000 times as long to obtain, the log-likelihood is not smooth enough for the optimization to converge.

As the choices of $K$ and $M$ will affect the shape of $l^{(K)}(\cdot)$ and the Monte Carlo variability which we estimate using $\sigma^2$, we consider two combinations of these values for each model. 
This practical choice is necessary for any SMC-based estimation, regardless of the strategy used to search $\boldsymbol\Theta$.
As the difference between $l(\cdot)$ and $l^{(K)}(\cdot)$ is of order $1/K$ \citep{bally1995law}, $K$ should be selected to be large enough to reduce this bias to some acceptable level, while keeping computational cost in mind.
Once $K$ has been selected, we choose $M=K^2$, which is the most efficient use of computational resources for the modified Brownian bridge sampler \citep{StramerYan2007b}.
We note that when $K$ and $M$ are specified to be large, it is even more important to carefully explore the parameter space, as each $Y(\cdot)$ value is more difficult to obtain. 

In the following simulations we choose $\mu_L\left(\btheta\right)=\beta$ and $R(\boldsymbol\theta,\boldsymbol\theta^\prime)=\exp\Big(-{||\boldsymbol\theta-\boldsymbol\theta^\prime||^2}/{\eta}\Big)$. We also assume $[\beta,\zeta]\propto{\eta}/{(\sigma^2+\tau^2)}$ where $\beta\in\mathbb{R}$ and $\sigma^2,\tau^2,\eta>0$. 
We find that even when these choices do not reflect reality, the sequential kriging-based optimization performs well.
We used R \citep{Rcite} for the implementation of SKBO and sped up the matrix calculations involved in calculating the expected improvement using C++ through the R package \textit{RcppArmadillo} \citep{RcppArmadillocite}.
We also used C to quickly obtain the modified Brownian bridge sample based on an adaptation of code in the R package \textit{sde} \citep{sdecite}, and the \textit{latin.hypercube} function of the R package \textit{emulator} \citep{emulatorcite} to obtain the initial grid over $\boldsymbol\Theta$.

\subsection{Ornstein-Uhlenbeck Process}
We first consider the Ornstein-Uhlenbeck (OU) model \citep{uhlenbeck1930theory} given~by
\begin{align}
\label{OUModel}
dX_t= \left(\theta_0+\theta_1 X_t\right)\,dt+dW_t,\quad 0\le t \le T,
\end{align}
where $X_0=x_0$ is the initial value of the process, $\theta_0\in\mathbb{R},\theta_1<0,$ and $W_t$ is a standard Brownian motion.
Without loss of generality, we have assumed unit diffusion in \eqref{OUModel}.
If this is not the case, \eqref{OUModel} may be transformed into a process with unit diffusion via the Lamperti transform \cite[see][for example]{ iacus2009simulation}.
This process has a transition density that is available in closed-form, so we can use the true log-likelihood and MLE to evaluate the performance of SKBO and the naive method.
We obtain our simulated data by first simulating $X_{0}$ from the stationary distribution of \eqref{OUModel}, given~by
$p(X_0,\btheta)= 
\phi\left(
X_0;\;
{\theta_0}/{\theta_1},\;
{1}/{\sqrt{2\theta_1}}
\right   ).
$
We then sequentially simulate $X_{t_1},\dots,X_{t_N}$ from the conditional distribution of \eqref{OUModel}, which is given~by
\begin{align}
\label{OUCondDist}
p(X_\Delta|X_0,\btheta)= 
\phi\left(
X_\Delta;
X_0 e^{-\theta_1 (1-\Delta)}+\frac{\theta_0}{\theta_1}\left[1-e^{-\theta_1(1-\Delta)}\right],
\frac{1}{2\theta_1}\left[1-e^{-2\theta_1(1-\Delta)}\right]
\right   ),
\end{align}
where $N$ is chosen to be 1000 and $t_i-t_{i-1}=0.1$, for $i=1,\dots,N$.
We use $\theta_0=2$ and $\theta_1=-3$ for this analysis. 
After repeating the analysis for various combinations of $\theta_0$ and $\theta_1$, we found the results presented below to be insensitive to the choice of these parameters.

\begin{figure}[t]
\centering
\includegraphics{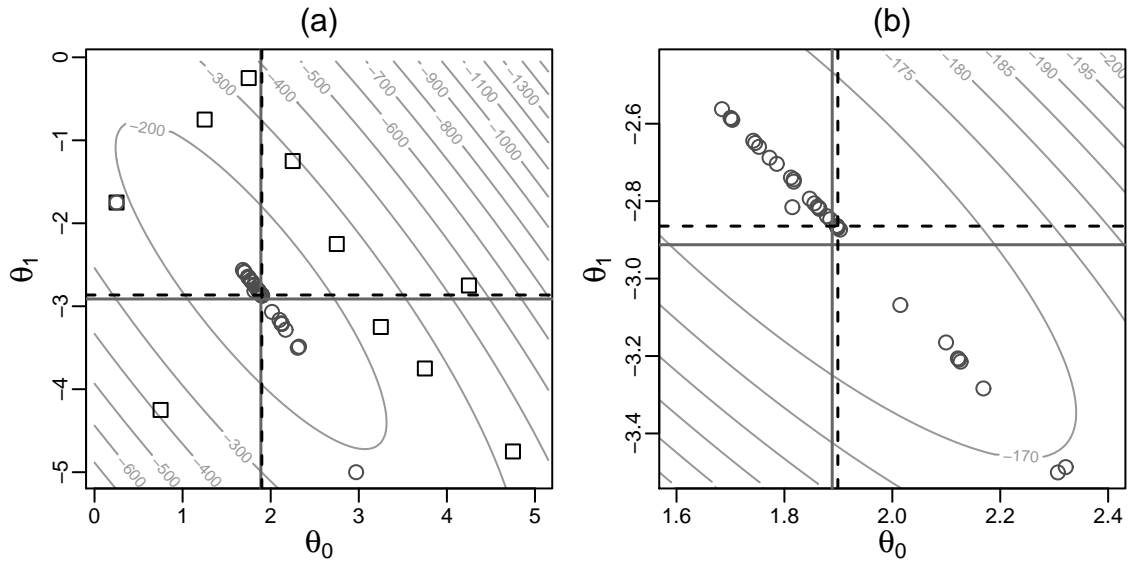}
\caption{
(a) A contour plot of the discretized log-likelihood for the OU
process given by \eqref{OUModel} with $\theta_0 = 2$ and $\theta_1 = -3$.  The
solid horizontal and vertical lines denote the exact MLEs of
$\theta_0$ and $\theta_1$ respectively, and the dashed lines denote
the SKBO-based estimate.  For the SKBO method, the squares indicate
the initial parameter values, and the circles denote the values added
sequentially.   (b) is a zoomed in version of (a).}
\label{SearchPath}
\end{figure}

We first investigate how well the SKBO method finds the maximum for
one simulated realization from \eqref{OUModel}.  Panels (a) and (b) of
Figure \ref{SearchPath} displays, as a contour plot, the discretized
log-likelihood for the realization of the OU process (Panel (b) is a
zoomed in version of (a)).  Looking at the figure, we can see that the
likelihood is concentrated along a line of positive slope in the
$(\theta_0, \theta_1)$ space.
In Figure \ref{SearchPath}, the squares denote the initial sample of
10 points generated using a Latin hypercube and the circles denote the
$\vv{\theta}$ values added using SKBO method, maximizing the expected
improvement as each new parameter value is added.  The majority of the
circles are concentrated along a ridge of high log-likelihood in the
parameter space, which indicates that the method is able to isolate an
estimate of the MLE.  A few circles further away from the ridge have
been added to decrease the uncertainty in estimating the discretized
log-likelihood using the GP.  On the figure the horizontal and
vertical lines denote the exact MLEs (solid line) and SBO-based
estimates (dashed line) of $\theta_0$ and $\theta_1$, respectively.
The estimates are close for this realization.

\begin{table}[t]
\centering
\caption{A comparison of the true MLE for 1000 simulations of an OU process given by
  \eqref{OUModel} with the MLEs obtained using two SMC-based methods:
  the ``Naive'' space-filling method and the ``SKBO'' method.  For
  each SMC-based method we vary $K$ and $M$ to control the accuracy of the SMC approximation. 
  We also vary the number of points sampled in the parameter space.
  The bootstrap standard error for the bias, SD, and RMSE is bounded above by $0.02$.}
\vspace{.25cm}
\begin{tabular}{|cc|c|cc|cc|cc|cc|}
\hline
           &      &     & \multicolumn{4}{|c|}{$K=5$, $M=25$} &   \multicolumn{4}{c|}{$K=10$, $M=100$}\\
\cline{4-11}
           &      & MLE & \multicolumn{2}{|c|}{Naive} & \multicolumn{2}{c|}{SKBO} & \multicolumn{2}{c|}{Naive} & \multicolumn{2}{c|}{SKBO} \\
\hline
\multicolumn{2}{|r|}{Initial pts}   & -- & 50 & 2500 & 10   & 20   & 50 & 2500 & 10   & 20   \\
\multicolumn{2}{|r|}{Avg added} & -- & -- & --   & 24.9 & 13.4 & -- &  --   & 21.7 & 10.6 \\
\hline
& Bias & 0.03
& 0.08 & 0.08
& 0.07 & 0.07
  & 0.05 
& 0.06 & 0.06
& 0.07 \\
$\theta_0$ & SD 
& 0.23 & 0.56
& 0.42 & 0.32
& 0.28 & 0.48
& 0.36 & 0.27
& 0.24 \\
& RMSE
& 0.22 & 0.56
& 0.43 & 0.33
& 0.29 & 0.48 
& 0.36 & 0.27 
& 0.24 \\
\hline
& Bias & 0.06
& 0.12 & 0.13
& 0.07 & 0.05
& 0.08 & 0.09 
& 0.05 & 0.07 
\\
$\theta_1$ & SD 
& 0.28 & 0.66
& 0.55 & 0.42
& 0.36 & 0.61 
& 0.47 & 0.35
& 0.31 \\
& RMSE
& 0.29 & 0.67
& 0.57 & 0.43
& 0.36 & 0.61 
& 0.48 & 0.35 
& 0.32 \\
\hline
\multicolumn{2}{|r|}{Coverage}   & 98.1\% & -- & -- & 75.6\% & 84.5\% & -- & -- & 82.9\% & 88.6\% \\
\hline
\multicolumn{2}{|r|}{Avg Time}
& -- & 20.5 & 1030.7 & 20.8 & 18.6 & 164.9 & 8262.7 & 122.1 & 105.6 \\
\hline
\end{tabular}
\label{OUTable}
\end{table}

Now we evaluate the general performance of SKBO-based estimation of
$\vv{\theta}$ relative to the exact MLE and naive space-filling
methods.  Table \ref{OUTable} compares the performance of each method
as we vary the accuracy of the SMC approximation for the naive and
SKBO methods (controlled by $K$ and $M$ -- we compare $K=5$
with the $K=10$ case) and as we change the
initial number of points sampled in the parameter space.  For this
simple process with $p=2$ parameters we are able to add a more
computationally intensive design which samples $2500$ points, a number
that could prove to be computationally impractical for other
processes.

In the table a number of different criteria are compared.  To compare
the quality of the estimates we summarize the bias, standard deviation
(SD), and root mean square error (RMSE) for each method. The bootstrap
standard error for the bias, SD, and RMSE is bounded above by 0.02. In
addition to recording the average time to find the approximate MLE,
and the number of points used initially for the SMC-based methods, we
present the average number of points added for the SKBO method.

Comparing to the gold standard of the exact MLE, as expected, the
approximate SMC-based methods (naive and SKBO methods) have a larger
bias, SD, and RMSE. Accounting for the uncertainty of these measures,
the SKBO method outperforms the naive space-filling methods,
regardless of the choice of the number of initial points.  Compared to
the SKBO method the 2500-point naive method still has disappointing
performance.  Also, the SKBO method takes much less time and uses
significantly less points (at least 32\% less points on average) to
get a better estimate than the naive space-filling method.  In terms
of obtaining a better estimate of the MLE using the SKBO method, we do
better for a larger initial number of points, and we improve as the
accuracy of the SMC approximation improves.  Note that as we increase
the approximation accuracy we need slightly less parameter values in
total.

We also compared the coverage of 95\% confidence regions for
$\vv{\theta}$ using the exact MLE with our SKBO-based method.  The
results are presented in the last line of Table \ref{OUTable}.  With
1000 replicates, testing that the coverage is equal to 95\%, a
$\alpha=0.05$ rejection region for coverages is below 96.3\% and above
96.4\%.  Thus the coverage for the exact MLE of 98.1\% is too high.
The SKBO-based coverages are too low, but the coverages increase to
88.6\% as we increase the accuracy of the SMC approximation and we
increase the initial number of points used.
%
%Additional calculations
%indicated that we were closer to the nominal coverage for 30 initial
%points with $K=10$ and $M=100$.
%
The accuracy of the SKBO-based estimate of the discretized
log-likelihood will depend on the degree of difference between the
specified GP model and the true log-likelihood. By focusing on the
log-likelihood near the maximum, the shape of the confidence regions
based on SKBO will be heavily influenced by the shape of the
log-likelihood near the estimated peak. To the degree that the
behavior of the log-likelihood near the maximum is not reflective of
its behavior elsewhere, the confidence regions will be
anti-conservative.  Increasing the number of initial points alleviates
this problem.

\subsection{Generalized CIR Model}
We now consider the generalized Cox-Ingersoll-Ross (GCIR) model,
introduced in \cite{chan1992empirical}, and analyzed in
\cite{RobertsStramer}.  The process is defined by
\begin{align}
\label{GCIRModel}
d{X}_t= \left(\theta_0+\theta_1{X}_t\right)\,dt+\gamma {X}_t^\psi\,dW_t,\quad 0\le t \le T,
\end{align}
where $X_0=x_0$ is the initial value of the process,
$\theta_0,\theta_1\in\mathbb{R},\gamma>0,\psi\in[0,1]$, and $W_t$ is a
standard Brownian motion.
We note that \eqref{GCIRModel} does not have a closed-form likelihood,
except for when $\psi = 0$ (the OU process) or $\psi = 0.5$ (the
Cox-Ingersoll-Ross model).
To improve our exploration of the parameter space, we let
$\theta_2=\log(\gamma)$ and $\theta_3=\log({\psi}/{(1-\psi)})$, and
optimize over the real-valued parameters $\vv{\theta} = (\theta_0,
\theta_1, \theta_2, \theta_3)$.

To simulate from the GCIR process we first simulate 100000 data points at
time increments of 0.001 from \eqref{GCIRModel} based on the Euler
approximation with $\theta_0=0.5$, $\theta_1=-0.25$, $\gamma=1$, and
$\psi=0.75$.  Next, subsampling every 100 time points, we obtain a
realization of length $N=1000$ and sampling interval $\Delta=0.1$. 

\begin{table}[t]
\centering
\caption{A comparison of the estimate MLEs from 1000 simulations of a GCIR process given by
  \eqref{GCIRModel} with the MLEs obtained using two SMC-based methods:
  the ``Naive'' space-filling method and the ``SKBO'' method.  For
  each SMC-based method we vary $K$ and $M$ to control the accuracy of the SMC approximation. 
  We also vary the number of points sampled in the parameter space.
  The bootstrap standard error for the bias, SD, and RMSE is bounded above by $0.03$.}
\vspace{.25cm}
\begin{tabular}{|cc|c|cc|c|cc|}
\hline
           &      &     \multicolumn{3}{|c|}{$K=5$, $M=25$} &   \multicolumn{3}{c|}{$K=10$, $M=100$}\\
\cline{3-8}
           &      & \multicolumn{1}{|c|}{Naive} & \multicolumn{2}{c|}{SKBO} & \multicolumn{1}{c|}{Naive} & \multicolumn{2}{c|}{SKBO} \\
\hline
\multicolumn{2}{|r|}{Initial pts}   & 100 & 20   & 40  & 100  & 20   & 40   \\
\multicolumn{2}{|r|}{Avg added} & --	& 60.3 		& 47.0		& -		& 56.2		& 46.3 \\
\hline
	 	& Bias 		& 0.21  	& 0.16 		& 0.12	 	& 0.27		& 0.11		& 0.13 		\\
$\theta_0$	& SD 		& 0.77 		& 0.54	 	& 0.64		& 0.79		& 0.52		& 0.49		\\
	 	& RMSE	& 0.79 		& 0.56 		& 0.65 		& 0.83		& 0.54		& 0.50		\\
\hline
	 	& Bias 		& -0.06	& -0.10 	& -0.05	& -0.10	& -0.08	& -0.11		\\
$\theta_1$	& SD 		& 1.11		& 0.80 		& 0.85		& 1.05		& 0.72		& 0.76		\\
	 	& RMSE	& 1.11		& 0.81 		& 0.85		& 1.06		& 0.72		& 0.77		\\
\hline
	 	& Bias 		& 0.21		& 0.19 		& 0.22 		& 0.16		& 0.12		& 0.13		\\
$\theta_2$	& SD 		& 0.17		& 0.18 		& 0.19 		& 0.18		& 0.12		& 0.11		\\
	 	& RMSE	& 0.27		& 0.26 		& 0.29 		& 0.24		& 0.17		& 0.17		\\
\hline
	 	& Bias 		& 0.21		& 0.25 		& 0.18		& 0.18		& 0.22		& 0.22 		\\
$\theta_3$	& SD 		& 0.83 		& 0.75 		& 0.79		& 0.87		& 0.79		& 0.71		\\
	 	& RMSE	& 0.85 		& 0.80 		& 0.81 		& 0.89		& 0.82		& 0.74		\\
\hline
\multicolumn{2}{|r|}{Avg Time} & 176.4	& 183.7 	&180.1  	& 1430.0 	& 1120.5	& 1245.5	\\
\hline
\end{tabular}
\label{GCIRTable2}
\end{table}

Table \ref{GCIRTable2} compares the naive and SKBO methods for
approximating the MLEs for the GCIR process.   The format of the table
is similar to that of Table \ref{OUTable}, with the exception that we
are not able to calculate an exact MLE.
The table illustrates that the naive method and SKBO perform similarly with respect
to estimating $\theta_2$ and $\theta_3$ when $K=5$, but the SKBO
method appears to slightly outperform the naive method when $K=10$ by these measures.
For estimating $\theta_0$ and $\theta_1$, similar to the OU case, SKBO
noticeably outperforms the naive method, regardless of the
accuracy of the SMC approximation.
%
%It appears that $\theta_2$ is the most important parameter in
%determining the shape of the log-likelihood, and thus both methods
%estimate this parameter better than the other three.
%
It is difficult to distinguish between the performance of the SKBO
method on the basis of the number of initial points used.  With a
smaller initial design we evaluate less points on average, but the timing,
bias, SD, and RMSE values are harder to discriminate.
This suggests that the performance of the SKBO method can be improved
by increasing the initial number of points up to some fixed number,
but that there is relatively small benefit in increasing the initial
number of points past this point.
Overall we stress that this number (roughly 20 in both the OU and GCIR
cases) is very small relative to the number of points required for the
naive space-filling method.

\section{An application to Modeling Stock Prices}
\label{sec:data}

\begin{figure}[t]
\centering
\includegraphics{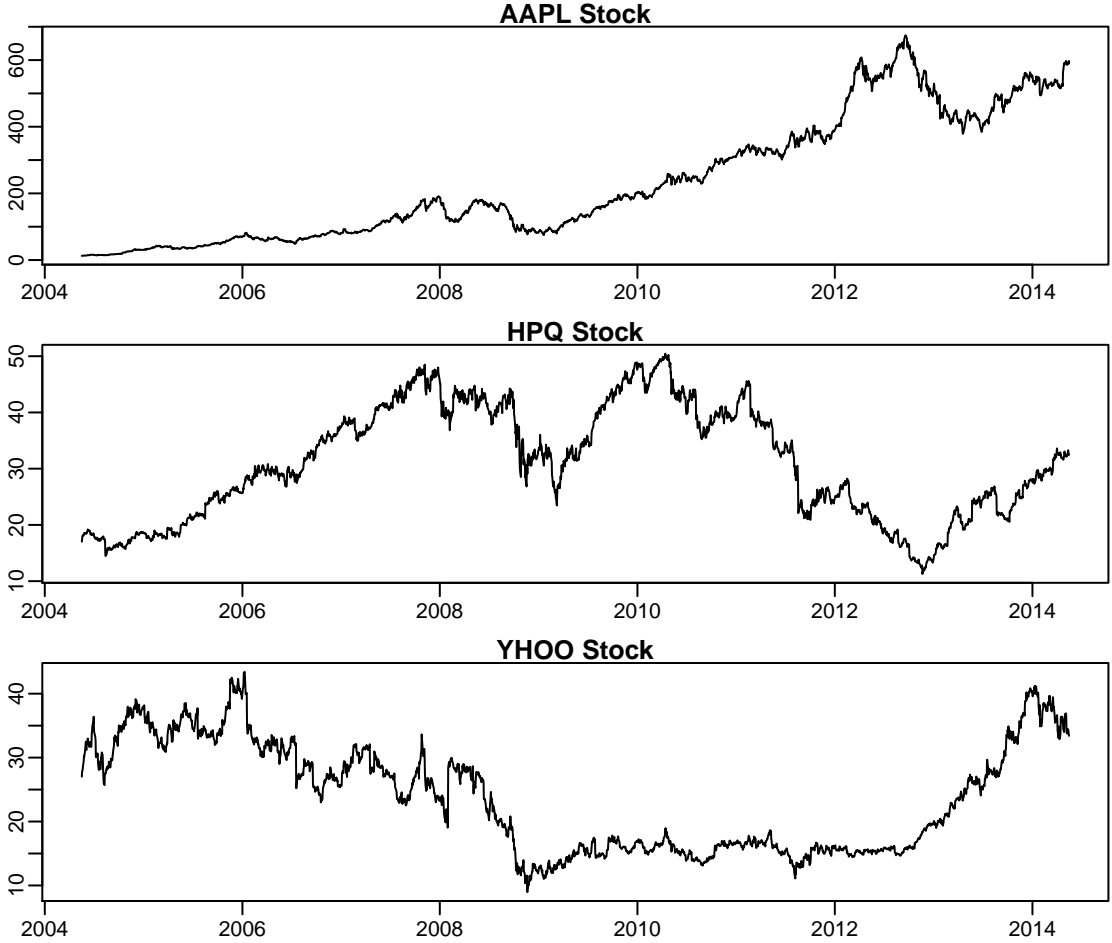}
\caption{Daily adjusted closing price for Apple, Inc., Hewlett-Packard Co., and Yahoo! Inc.}
\label{DailyStockPrice}
\end{figure}

In this section we apply the methodology described above to modeling
the stock price of three technology companies: Apple, Inc.~(AAPL),
Hewlett-Packard Co.~(HPQ), and Yahoo! Inc.~(YHOO).
We consider the daily adjusted closing price of each stock from the
ten-year period of May 17th, 2004 through May 16th, 2014 (2518
observations for each series).
The data, shown graphically in Figure \ref{DailyStockPrice}, were
obtained from Yahoo! Finance (\url{http://finance.yahoo.com}).  All
three series exhibit drift, local trends, and volatility.

An popular model for stock price is the \emph{geometric Brownian
  motion} (GBM) process, which is the solution to
\begin{align}
dX_t=\theta_0 X_t\,dt+ \gamma X_t\, dW_t, \quad 0\le t \le T,
\label{GBMModel}
\end{align}
where $X_0=x_0$ is the initial closing price of the process,
$\theta_0\in\mathbb{R}, \gamma>0$, and $\{W_t\}$ is a standard
Brownian motion.
Stock prices were modeled by \eqref{GBMModel} in the famous
Black-Scholes model \citep{black1973pricing} for option pricing and
\eqref{GBMModel} is refered as ``the model for stock prices'' in
\cite{hull2012options}, a popular introductory finance text.
We choose a sampling interval of $\Delta=1/252$ as there are roughly
252 trading days per year.
As the SDE described by \eqref{GBMModel} has a known transition
density, we can compute the MLEs and the corresponding log-likelihood
for each stock.
These values are displayed in the columns of Table~\ref{GBMResults}
with the heading ``GBM''.

\begin{table}[t]
\centering
\caption{Maximum likelihood estimates, maximized log-likelihoods, and
  AIC for the geometric Brownian motion and generalized geometric Brownian
  motion models for three difference daily adjusted closing stock
  prices from May 17th, 2004 through May 16th, 2004. The likelihood
  ratio statistic compares the generalized to the non-generalized model.}
\vspace{.25cm}
\begin{tabular}{|c|cccc|ccccc|c|}
\hline
&\multicolumn{4}{|c|}{GBM}&\multicolumn{5}{|c|}{Generalized GBM}&\\
Stock	& $\widehat{\theta}_0$ & $\widehat{\gamma}$ & Log-lik. & AIC &
$\widehat{\theta}_0$ & $\widehat{\gamma}$ & $\widehat{\psi} $ &
Log-lik. & AIC & LRT stat.\\
\hline
AAPL 	& 0.45	& 0.37	& -6797 	& 13598	& 0.34	& 1.09	& 0.79	& -6740	& 13486 	& 114	\\  
HPQ 	& 0.12	& 0.33	& -2341 	& 4686	& 0.30	& 1.27	& 0.61	& -2331	& 4668	& 20	\\
YHOO	& 0.10	& 0.40	& -2149 	& 4302	& 0.37	& 1.03	& 0.71	& -2135	& 4276	& 28	\\
\hline
\end{tabular}
\label{GBMResults}
\end{table}

As a possible alternative model, we consider the \emph{generalized}
GBM given by the solution to
\begin{align}
dX_t=\theta_0 X_t\,dt+ \gamma X_t^\psi\, dW_t, \quad 0\le t \le T,
\label{GenGBMModel}
\end{align}
where $X_0=x_0$ is the initial closing price, $\theta_0\in\mathbb{R},
\gamma>0, \psi\in[0,1]$, and $\{W_t\}$ is a standard Brownian motion.
That is, we investigate whether the model for these three stock prices
can be improved by allowing $\psi$ to differ from one.
As \eqref{GenGBMModel} does not have a known transition density, we
estimate the MLEs for each stock using SKBO with $K=10$ and $M=100$.
Once these estimated MLEs have been obtained, we estimate the
log-likelihood using the modified Brownian bridge sampler with $K=20$
and $M=400$.
This highlights the important point that while the estimated MLE may
be obtained for fairly small values of $K$ and $M$, estimating the
log-likelihood accurately will typically require larger values.
Also, while the kriging mean $\eta_L(\cdot)$ is very useful for
guiding the search of $\boldsymbol\Theta$, we do not use it to
estimate the maximized log-likelihood as this would essentially be
underestimating the maximum using a mean.
The parameter estimates and (estimated) maximized likelihood values
are displayed in the columns of Table~\ref{GBMResults} with the heading
``Generalized GBM''.

The log-likelihood for the generalized GBM model is higher than that
modeled for the GBM for each of the three stock series.
Evaluating the two models based on the Akaike Information Criterion
(AIC), after accounting for an extra parameter, the
generalized GBM model outperforms the usual GBM model for each of
these three stocks.
Using the fact that the models are nested, we can also compute the
likelihood ratio statistic (LRT) to test the null hypothesis that the
GBM model fits sufficiently well to each stock prices series versus
the alternative hypothesis that the generalized GBM is required.
The LRT statistics given in the last column of Table~\ref{GBMResults}
confirm that the generalized model is more reasonable for all three
series.  We conclude that the future volatility in all of these
technology stocks is dependent on the discounted current stock price.

\section{Discussion and Future Work}
\label{sec:discussion}

In this research we introduce a sequential, kriging-based optimization
strategy which provides a derivative-free method for approximating the
MLE in SDE models, in cases where the likelihood function cannot be
evaluated exactly, but can be estimated with some degree of
statistical accuracy. Our work is primarily motivated by the case
where the data are discrete-time observations of an SDE; however, what
we suggest can be extended to other settings without difficulty. We
judge the performance of our approach on two fronts: (1) statistical
accuracy, as measured by the bias, SD, and RMSE of the estimated MLE,
and (2) computational efficiency. Our findings show that the proposed
method outperforms space-filling competitors, even in the cases
where these methods use significantly more likelihood evaluations.

A significant component of the SKBO method is the assumption that the
unobserved likelihood is a realization of a certain GP, specified via
its mean and covariance structure. The validity of this assumption is
difficult to establish. However, in the cases where we do have access
to the exact likelihood function we do not find any evidence to
disprove it.  On the other hand, the Gaussian assumption offers the
significant advantage of yielding a closed form expression for the
expected improvement. In addition, the GP-method allows us to provide
measure of uncertainty for our estimates (the MLE and the underlying
estimates of the likelihood function). For the examples discussed
above, we use simple and popular parametric functions to describe this
underlying GP. We show that even a naive formulation leads to
impressive results. We do however acknowledge that the
parameterization of the GP process may be crucial in certain
applications. For examples with a more involved log-likelihood
function, a non-stationary GP process may be necessary to improve the
estimation of the MLE. With more sophisticated GP processes, we would
need to sample more parameter values to learn about the unknown
log-likelihood function and its inherent uncertainty. We take a
Bayesian viewpoint to learn and update the GP parameters, thus
naturally incorporating uncertainty and allowing us to take advantage
of prior knowledge.

From a computational perspective, we identify two bottlenecks: (1)
maximizing the expected improvement and (2) updating the structure of
the underlying GP given that a new parameter has been added to the
procedure. For a specific application one can use state of the art
software to reduce the computational overhead in each case.  In the
applications presented in this paper we implement the SKBO method
using multi-core R routines. This is done in order to allow for a fair
comparison to other approaches. Our method lends itself naturally to
parallel computing, given that we make heavy use of Monte Carlo and
importance sampling techniques. Thus, we expect that a more
sophisticated approach which uses several CPUs or even a Graphical
Processing Unit may offer significant improvements.

Modern statistical applications involve observational models with ever
increasing complexity. In many cases, an exact evaluation of the
likelihood function is impossible and thus one is forced to use an
array of approximations. Our proposed method offers several
advantages, making it a valuable addition to the MLE toolbox. We are
currently investigating its application to the analysis of other
statistical models.

% ======================================================================
\paragraph{Acknowledgments.}
Thank you to Professor Thomas Santner for constructive discussions on
this work. Herbei is supported in part by the US National Science
Foundation under grant DMS-1209142.

\appendix
\section{Appendix A}
%{Modified Brownian Bridge}
\label{sec:modBB}
As noted in \cite{StramerYan2007b}, $R^{(K)}$ may have unbounded
variance, causing the sampler to be inefficient. A common claim in the
existing literature is that increasing $K$ will decrease the
discretization error (bias) and increase the Monte Carlo error
(variability) while increasing $M$ will reduce the
variability. However, we have found that for any $M$, increasing $K$
past some optimal choice actually \emph{increases} the bias in the
estimated likelihood. This problem grows worse as $N$, the amount of
data, increases. In \cite{brandt2002simulated}, the convergence of the
simulated MLE to the true MLE is shown when $K\rightarrow\infty$,
$M\rightarrow\infty$, and $\sqrt{M}/K\rightarrow 0$. The true MLE in
turn converges to the true parameter vector as
$N\rightarrow\infty$. From our investigation, the likelihood is most
biased when $\sqrt{M}/K$ is small and $N$ is large (which may or may
not be reflected in the simulated MLE).

\cite{DurhamGallant2002} propose two Monte Carlo samplers based on
Brownian bridges.  The Brownian bridge sampler is defined to be the
Euler approximation of the solution $H_t$ to
\begin{align}
\label{BBridgeSamp}
dH_t=\frac{X_\Delta-H_t}{\Delta-t}dt+\sigma(H_t,\boldsymbol\theta)dW_t, & \mbox{  } & 0\le t\le\Delta, & \mbox{  } & H_0=X_0.
\end{align}
For a constant $\sigma(\cdot)$, $H_t$ is a Brownian bridge on
$[0,\Delta]$ from $X_0$ to $X_\Delta$.

The modified Brownian bridge sampler is introduced based on the recursion
\begin{align}
\label{AppmodBB}
X_{\tau_k}=X_{\tau_{k-1}}+\frac{X_{\tau_K}-X_{\tau_{k-1}}}{K-k+1}+\sqrt{\frac{K-k}{K-k+1}}\sqrt{\frac{\Delta}{K}}\sigma(X_{\tau_{k-1}},\boldsymbol\theta)Z_{k},& \mbox{ for } k=1,\dots,K-1,
\end{align}
where $X_{\tau_0}=X_0$, $X_{\tau_K}=X_\Delta$, and
$Z_{k}\overset{IID}\sim N(0,1)$. Note that this sampler is identical
to \eqref{BBridgeSamp} except for the ${(K-k)}/{(K-k+1)}$ term in the
variance. The authors state ``it is not entirely obvious that this
should be the case, [but] we will see that this modification results
in much better performance''. To lend support to the heuristic
arguments justifying \eqref{AppmodBB}, \cite{StramerYan2007} show that
when $\sigma(\cdot)$ is constant, the modified Brownian bridge is also
exactly a Brownian bridge on $[0,\Delta]$ from $X_0$ to $X_\Delta$. As
discussed in \cite{chib2002numerical}, $R^{(K)}$ based on this scheme
can be interpreted as a simple Euler approximation multiplied by the
expected value of the ratio of two predictive densities.

% ======================================================================
\setlength{\bibsep}{2pt}
\bibliographystyle{apalike}
\bibliography{References}
% ======================================================================

% ======================================================================
\end{document}